# Drug-delivery Ca-Mg silicate scaffolds encapsulated in PLGA

A. Jadidi [a], E. Salahinejad [*,a], E. Sharifi [b], L. Tayebi [c]

[a] Faculty of Materials Science and Engineering, K. N. Toosi University of Technology, Tehran, Iran

[b] Department of Tissue Engineering and Biomaterials, School of Advanced Medical Sciences and Technologies, Hamadan University of Medical Sciences, Hamadan, Iran

[c] Department of Developmental Sciences, Marquette University School of Dentistry, Milwaukee, WI 53233, USA

**Abstract**

The aim of this work is to develop dual-functional scaffolds for bone tissue regeneration and local antibiotic delivery applications. In this respect, bioresorbable bredigite ($Ca_7MgSi_4O_{16}$) porous scaffolds were fabricated by a foam replica method, loaded with vancomycin hydrochloride and encapsulated in poly lactic-co-glycolic acid (PLGA) coatings. Field emission scanning electron microscopy, Archimedes porosimetry and Fourier-transform infrared spectroscopy were used to characterize the structure of the scaffolds. The drug delivery kinetics and cytocompatibility of the prepared scaffolds were also studied *in vitro*. The bare sample exhibited a burst release of vancomycin and low biocompatibility with respect to dental pulp stem cells based on the MTT assay due to the fast bioresorption of bredigite. While keeping the desirable characteristics of pores for tissue engineering, the biodegradable PLGA coatings modified the drug release kinetics, buffered physiological pH and hence improved the cell viability of the vancomycin-loaded scaffolds considerably.

* Corresponding Author: Email Address: <salahinejad@kntu.ac.ir>

1. Introduction

One of the major complications of bone tissue engineering after implantation is bacterial bone infection or osteomyelitis, which can cause severe pain in patients and demand a high cost of treatment (Hetrick and Schoenfisch, 2006; Lucke et al., 2003; Shirtliff and Mader, 2000). To overcome this problem, systematic drug delivery is a prevalent method with several disadvantages like the possibility of high systemic toxicity and longer hospitalization. In contrast, local drug delivery is a suitable alternative that lacks the aforementioned drawbacks. These devices provide a sustained concentration of drugs needed for the inhibition of pathogens at the intended site in a prolonged and controlled fashion with no or minimal drug concentration in the systemic circulation (Jain, 2008; Marwah et al., 2016; Sankar et al., 2011). Fortunately, tissue-engineering scaffolds have the ability to act as the matrix of local drug delivery systems for bone infection treatment at the implantation site. In this regard, chitosan-coated bioactive glass scaffolds loaded with gatifloxacin and fluconazole (Soundrapandian et al., 2010), gelatin-hydroxyapatite scaffolds impregnated with ciprofloxacin (Krishnan et al., 2015) and calcium sulfate scaffolds loaded with tobramycin (Ferguson et al., 2014) are noticeable.

In this study, bredigite ($Ca_7MgSi_4O_{16}$) was chosen to fabricate the foundation of a dual-functional system with the aim of bone tissue regeneration and local drug delivery. From the viewpoint of mechanical properties, it is well-established that Mg-Ca silicates including bredigite are superior to apatites and bioglasses which are commonly used for implantation (Schumacher et al., 2014; Wu et al., 2007). As well as the direct role of Ca in bone-related processes, the presence of Mg and Si in this group of bioceramics is beneficial for some cellular processes and human metabolism, including bone healing processes (Diba et al., 2014; Diba et al., 2012). In the orthorhombic structure of bredigite, the most abundant cation is $Ca^{2+}$, which provides high resorbability due to the lower activation energy of Ca-O bond for hydrolysis than Mg-O bond (Wu and Chang, 2007a). Among the variety of antibiotics, vancomycin was also selected to treat osteomyelitis because this is a very active glycopeptide antibiotic against *staphylococcus aureus*. In addition, this has the least deleterious effect on the function of osteoblasts and skeletal cells compared to other antibiotics used in the treatment of osteomyelitis like ciprofloxacin and tobramycin (Antoci Jr et al., 2007; Edin et al., 1996).

The adsorption of drugs on drug delivery devices with weak physical bonds like the van der Waals type, rather than a chemical interaction, gives rise to a burst release of the drug during implantation. This is disadvantageously accompanied by the release of the entire drug before controlling the infection and a detrimental effect on biocompatibility due to the high concentration of the drug around the surrounding tissue (Huang and Brazel, 2001). It is noticeable that in the treatment of osteomyelitis caused by *staphylococcus aureus*, vancomycin should be released continually through 4-6 weeks at concentrations exceeding the minimum inhibitory concentration (MIC, 0.75-2 µg/ml) and minimum bactericidal concentration (MBC, 8 µg/ml) (Coudron et al., 1987; Gold and Moellering Jr, 1996; Mason et al., 2009). MIC is the minimal concentration of a drug which prevents the visible growth of a bacterium and MBC is the lowest antibacterial density necessary to kill the bacterium. On the other hand, the high bioresorption rate of bredigite leads to metabolic alkalosis due to the high concentration of alkaline ionic ($Ca^{2+}$ and $Mg^{2+}$) around the surrounding tissue followed by the reduction of hydrogen ions ($H^+$) outside the normal range (Galla, 2000). The use of polymer coatings is a promising approach to control both the resorption rate of scaffolds and the drug release rate of these devices.

To the best of our knowledge, there are no reports in the literature on the use of bredigite scaffolds as the matrix of drugs, which is the subject of this work. In this regard, to control the burst release of vancomycin and high physiological pH caused by the fast bioresorption of bredigite, bredigite scaffolds were coated with poly lactic-co-glycolic acid (PLGA). PLGA is a biodegradable synthetic aliphatic polyester made from the monomers of polylactide (PLA) and polyglycolide (PGA) with successful consequences as coating for drug delivery and tissue-engineering applications (Bose et al., 2018; Jadidi and Salahinejad, 2020; Khojasteh et al., 2016; Maurmann et al., 2017; Olalde et al., 2013; Su et al., 2019; Zamboni et al., 2017). Alternatively, polycaprolactone (PCL) is another polymer which has been extensively used for biomedical applications due to suitable biodegradability and nontoxicity (Park et al., 2012). However, PCL with a considerable biostability is postulated to disadvantageously suppress the appropriate bioresorbability of bredigite. The use of PLGA is also hypothesized to buffer the metabolic alkalosis phenomenon caused by the fast bioresorption of bredigite (Jadidi and Salahinejad, 2020; Keihan et al., 2020) since the degradation of this biopolymer acidifies the physiological medium (Zhao et al., 2011; Zhao et al., 2012).

## 2. Experimental procedures

### 2.1. Materials

Tetraethyl orthosilicate (($C_2H_5O)_4Si$, TEOS, Merck, Germany, Purity>98%), magnesium nitrate hexahydrate ($Mg(NO_3)_2 \cdot 6H_2O$, Merck, Germany, Purity>98%), calcium nitrate tetrahydrate (Ca(NO$_3$)$_2$.4H$_2$O, Merck, Germany, Purity>98%), and nitric acid (HNO$_3$, 2 M, Merck, Germany) were used for the sol-gel synthesis of bredigite. The drug and coating materials were vancomycin hydrochloride (C$_{66}$H$_{75}$Cl$_2$N$_9$O$_{24}$, Sigma Aldrich) and PLGA (5004A, lactide/glycolide ratio of 50:50, molecular weight = 44 kDa, acid-terminated, acid number: min 3 mg KOH/g, Corbion, Netherlands), respectively.

## 2.2. Synthesis of powder used for scaffolding

A sol-gel technique similar to Refs. (Jadidi and Salahinejad, 2020; Wu and Chang, 2007b; Wu et al., 2007) was used for the synthesis of bredigite. For the preparation of 1.000 g bredigite powder, 1.000 ml TEOS, 0.651 ml distilled water and 0.361 ml nitric acid were stirred at room temperature for 30 min. Thereafter, 0.289 g magnesium nitrate hexahydrate and 1.864 g calcium nitrate tetrahydrate were added to the solution and again stirred for 5 h. The solution was then maintained at 60 °C for 1 day, dried at 120 °C for 2 days and calcined at 700 °C for 3 h.

## 2.3. Fabrication of bredigite scaffolds

In order to fabricate bredigite scaffolds, the sacrificial foam replication method was employed. For this purpose, the calcined powder was suspended in a sodium alginate aqueous solution (3% w/w) with the bredigite/sodium alginate solution mass ratio of 1:3 under sonication for 3 min. Polyurethane foam templates (density: 25 ppi, porosity >97%) were cut in a dimension ~10×10×10 mm$^3$, immersed in a glass beaker containing the bredigite slurry and then forced by a compressed air flow to eliminate the extra slurry. After drying at 60 °C for 12 h, the impregnated foams were heated at 300 °C for 1 h for the removal of the urethane struts and then at 1350 °C for 3 h for sintering at the heating rate of 3 °C/min. X-ray diffraction analyses have previously confirmed the successful synthesis of bredigite via this processing route (Jadidi and Salahinejad, 2020; Wu and Chang, 2007b; Wu et al., 2007).

### 2.4. Drug loading in scaffolds

For the drug loading into the scaffolds, vancomycin hydrochloride was dissolved in distilled water with the concentration of 0.31 mg/ml. The scaffolds were maintained in the vancomycin solution with the ratio of 5 mg/ml for 24 h and then dried in air for 24 h.

### 2.5. PLGA coating of scaffolds

For the PLGA coating of the vancomycin-loaded scaffolds, 5 and 10 % w/v solutions of PLGA and acetone were prepared at room temperature. Thereafter, the scaffolds were dipped into 10 ml of the PLGA solution for 3 sec and dried at 60 °C for 2 h for the solvent removal.

### 2.6. Structural characterization

The morphology of the samples was analyzed by field emission scanning electron microscopy (FESEM, MIRA3 TESCAN, Czech Republic, accelerating voltage= 15 kV) after gold sputtering. Porosimetry was also performed using the water Archimedes method, according to the following formula (Mirhadi et al., 2012):

$$P = [(W_1-W_2)/(W_1-W_3)] \times 100 \tag{1}$$

where $P$ is the pore percentage, $W_1$ and $W_2$ are the weight of the scaffolds before and after immersion in water, respectively, and $W_3$ is the weight of the suspended scaffolds.

Fourier-transform infrared spectroscopy (FTIR, Thermo Nicolet, Avatar, USA) was also conducted on the milled scaffolds in the spectral range of 3600-400 cm$^{-1}$ with the resolution of 4 cm$^{-1}$ in a moisture-free environment at 25 °C.

## 2.7. Drug release analysis

In order to determine the *in vitro* release profile of vancomycin, the bare and PLGA-coated bredigite scaffolds impregnated with vancomycin were immersed into a glass bottle containing 20 ml phosphate buffered saline (PBS) with the ratio of 4 mg/ml at pH of 7.4 and 37 °C. At each predetermined time point, 3 ml of the solution was taken out and replaced with 3 ml of the fresh PBS solution. The amount of vancomycin in the solution was investigated by a UV spectrophotometer (UV–1100, MAPADA INSTRUMENT, Shanghai, China) at the wavelength of 280 nm with three repetitions. To calculate the accumulative release percentage of the drug, the total amount of vancomycin loaded into the scaffolds was also measured after the scaffolds were completely degraded in the medium. In order to establish the calibration curve of measurements, five concentrations of vancomycin in the range of 1-20 μg/ml were chosen. According to the Beer-Lambert law (Thomas, 1996), the following calibration equation was found with the correlation coefficient of $R^2 = 0.998$ and used:

Absorbance (counts) = 1.9 × Concentration (mg/ml) (2)

## 2.8. Physiological pH evaluation

The variations of physiological pH due to exposure to the scaffolds were investigated by immersing them in the simulated body fluid (SBF) at 37 °C for 7 days using an electrolyte type pH meter (PHS-2C, Jingke Leici Co., Shanghai, China) with three repetitions. To do so, the scaffold/SBF ratio of 1:200 g/ml (Wu et al., 2006; Wu et al., 2011; Wu et al., 2010) was used.

## 2.9. Biocompatibility assessments

The MTT assay was conducted to assess the cytocompatibility of the prepared scaffolds of 3×3×3 mm$^3$ in size. The scaffolds were sterilized via exposure to UV radiation for 2 h. 3,500 human dental pulp stem cells obtained from National Cell Bank of Iran were seeded on each set of the scaffolds with three repeatations in a 96-well culture plate with the scaffold-free cells-containing medium as the control and were then incubated at 37°C under 5% $CO_2$ with Dulbecco`s modified Eagle media (DMEM) supplemented with 10% fetal bovine serum (FBS, Gibco, USA) for 1, 3 and 7 days. The cytotoxicity assessment was done by using 3-[4, 5-dimethylthiazol-2-y1]-2, 5 diphenyltetrazolium bromide (MTT, Sigma, USA) colorimetric assay based on the instructions of Ref. (Shahrouzifar et al., 2019). For this purpose, the culture medium was first removed, and then 100 μL of the MTT solution (5 mg/ml) was added to each well and incubated for 2 h at 37 °C under 5% $CO_2$. Then, the medium was removed, and formazan precipitates were dissolved in dimethyl sulfoxide (Sigma, USA). The optical density of viable cells was measured by a microplate reader (ChroMate-4300, FL, USA) at the wavelength of 545 nm. One-way analysis of variance with a statistical significance level less than 5% (p-value<0.05) was used to compare the cytocompatibility data.

The cell morphology on the optimal scaffold was also evaluated according to the protocol of Ref. (Shahrouzifar and Salahinejad, 2019). Briefly, 20,000 cells were seeded onto the sterilized scaffold and incubated in DMEM, 10% FBS, 100 IU/mL penicillin, and 100 IU/mL streptomycin. After 24 h of cell seeding, cells were fixed on the scaffold by 300 μl glutaraldehyde (2.5% in PBS) for 3 h at room temperature, then washed with PBS for 5 min and dehydrated in ascending ethanol series (30, 50, 70, 80, 90, and 100%). Finally, the sample was coated with a thin layer of gold and studied by electron microscopy.

## 3. Results and Discussion

### *3.1. Structural analyses*

The FESEM micrograph of the powder calcined at 700 °C used as the feedstock of scaffolding is shown in Figure 1a, revealing highly agglomerated nanoparticles of almost 25 nm in size. Figures 1b, 1c and 1d present the macrograph and micrograph of the scaffolds sintered at 1350 °C, consisting of irregular shaped particles of about 1-10 μm in size. The mean size of pores inside the struts is 3 μm, while the struts form pores of almost 200-1000 μm in size which are essential for bone tissue engineering. It would be worth mentioning that natural cancellous and cortical bones have 75-85% and 5-10% porosity levels with 300-600 and 10-50 μm sizes, respectively (Lee et al., 2012; Weiner and Wagner, 1998).

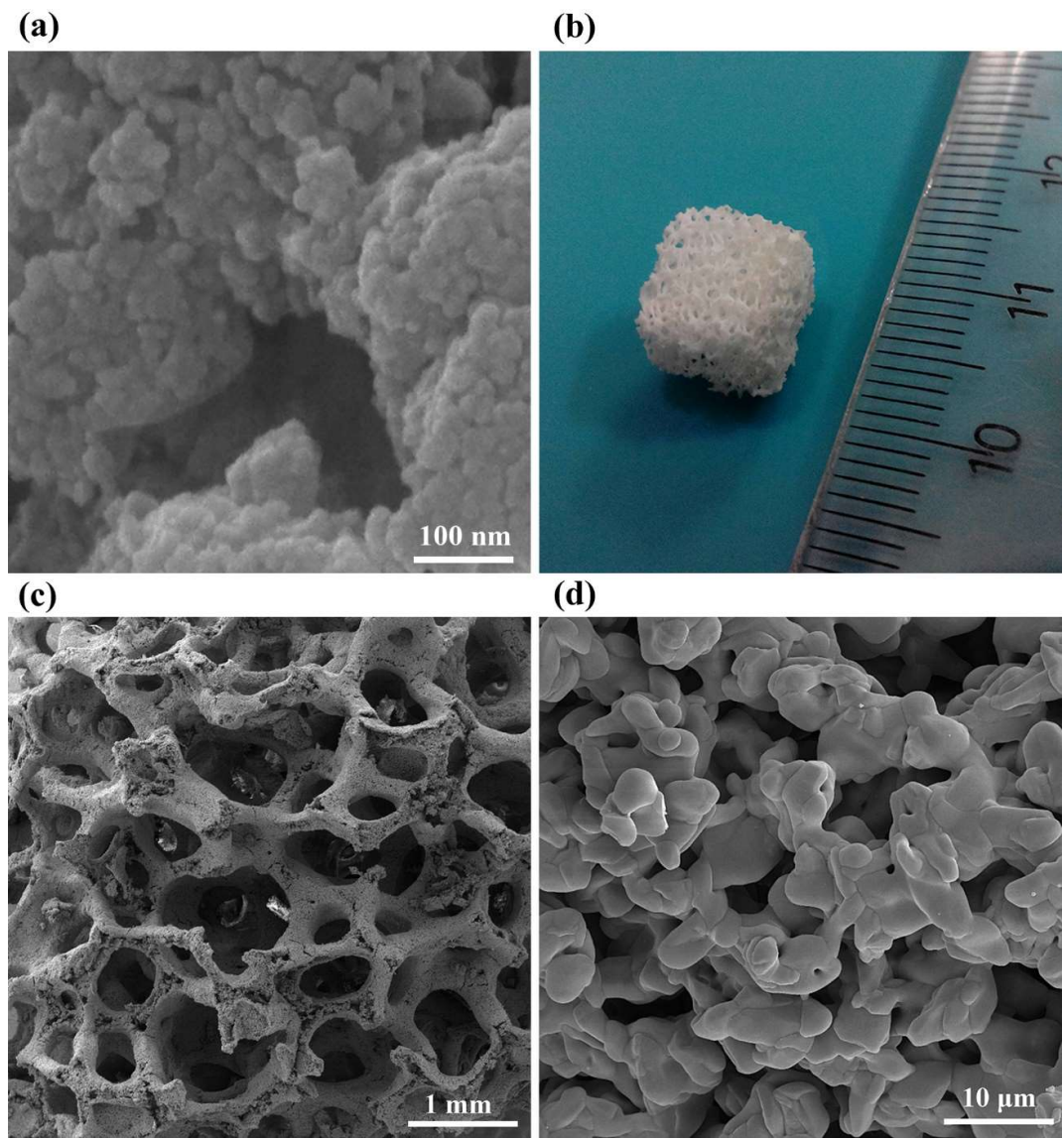

Figure 1. FESEM micrograph of the sol-gel derived powder calcined at 700 °C (a), macrograph of the scaffold sintered at 1350 °C (b) and FESEM micrographs of the sintered scaffold in two magnifications (c, d).

Figure 2 reveals the FESEM micrograph of the bare and PLGA-coated bredigite scaffolds. As well as the preservation of pores interconnectivity (Figures 2a, 2c and 2e), the filling of fine pores of the struts as the drug loading matrix is evident after PLGA coating (Figures 2d and 2f). Based on the Archimedes porosimetry method, the mean porosity levels of about 90, 82 and 77 % were also measured for the bare, bredigite-5% PLGA and bredigite-10% PLGA scaffolds, respectively, where all are in the desired range of tissue-engineering scaffolds. On the contrary, 15% PLGA coating has been reported to disadvantageously block the pores of the scaffold (Jadidi and Salahinejad, 2020); thus, this sample is not considered in this study. In addition, the thickness of the PLGA coatings deposited by the polymer concentrations of 5 and 10 % is measured to be almost 7 and 16 μm, respectively (Figure 3). It is also noticeable that the mean compressive mechanical strength of the bare, 5% PLGA-coated and 10% PLGA-coated bredigite scaffolds, albeit without any drug impregnation, is almost 0.3, 1.2 and 1.7 MPa, respectively (Jadidi and Salahinejad, 2020), where the strength of the PLGA-coated scaffolds is acceptable for bone tissue-engineering applications.

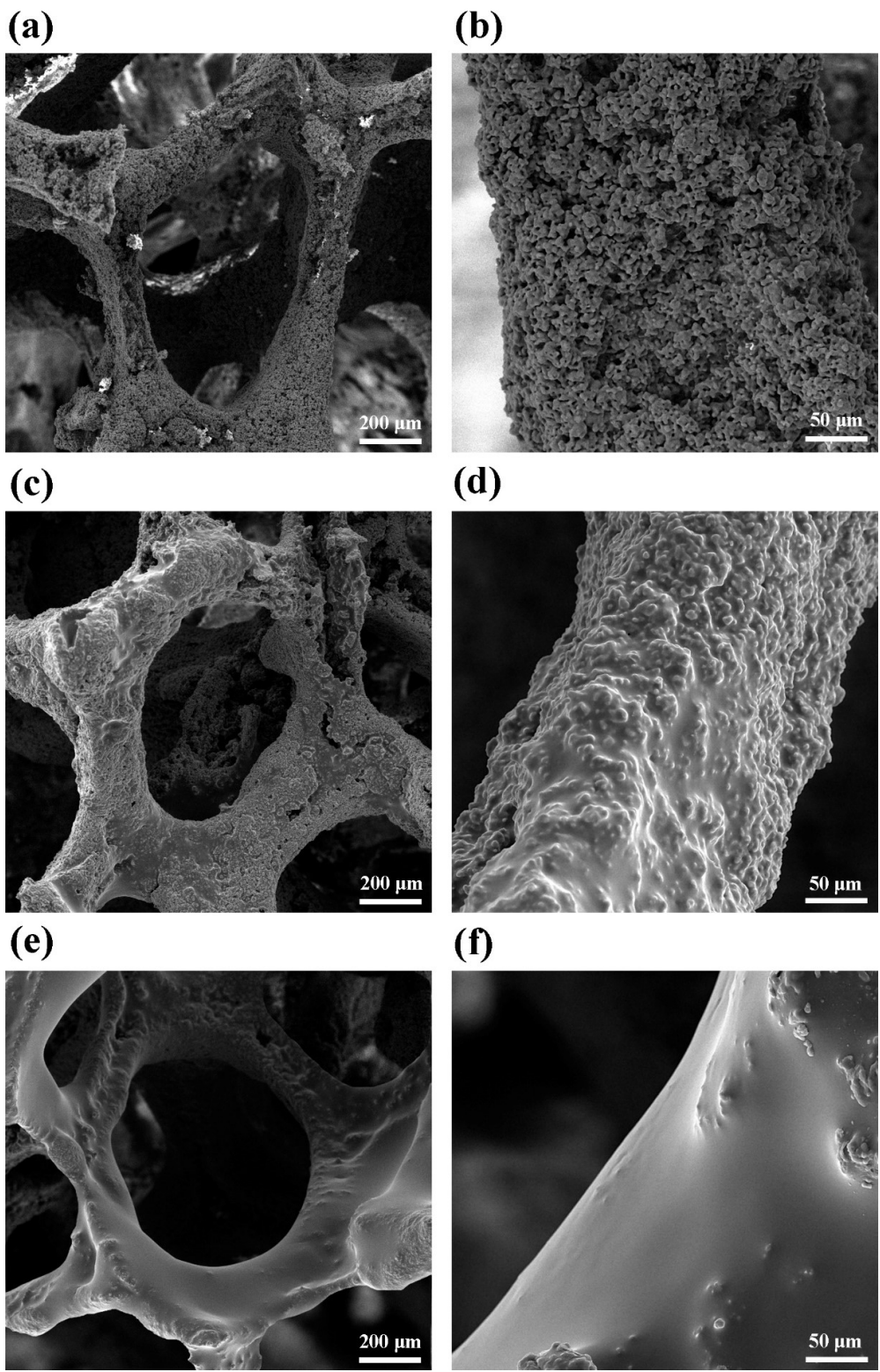

Figure 2. FESEM micrographs of the bare (a, b), 5% PLGA-coated (c, d) and 10% PLGA-coated (e, f) bredigite scaffolds in two magnifications.

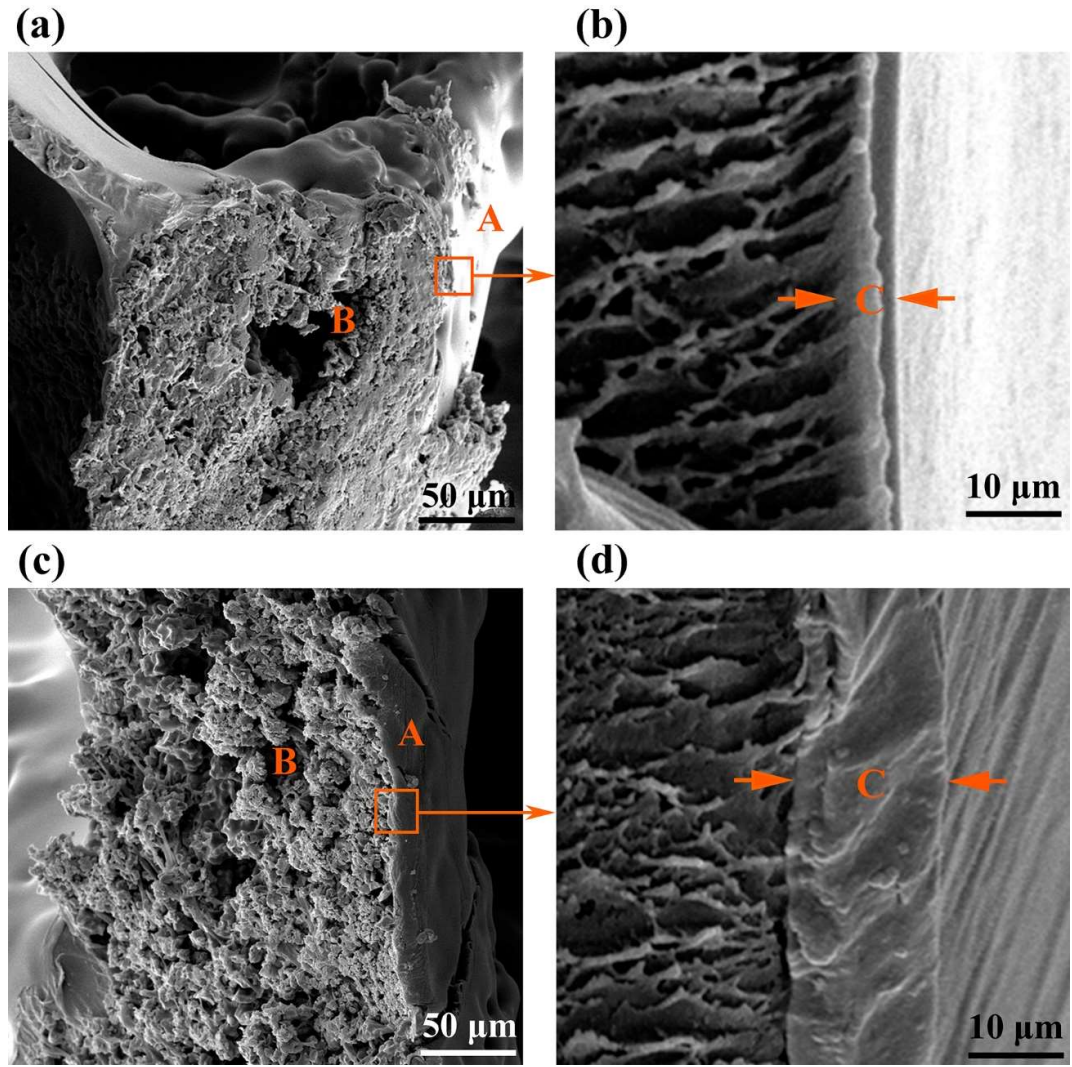

Figure 3. Cross-sectional FESEM micrographs of the 5% (a, b) and 10% (c, d) PLGA-coated bredigite struts, where A, B and C represent the PLGA coating surface, bredigite strut cross section and PLGA coating cross section, respectively.

In order to confirm the formation of bredigite, the incorporation of vancomycin and the deposition of PLGA, the FTIR spectroscopy on the samples was used (Figure 4). For the ceramic scaffold before impregnating with the drug and PLGA (Figure 4a), the characteristic peaks of 1100-1000, 960 and 873 cm$^{-1}$ are assigned to the SiO$_4$ tetrahedron stretching. The peak of 616 cm$^{-1}$ is related to the SiO$_4$ tetrahedron bending. Also, 516 and 553 cm$^{-1}$ are attributed to the O–Mg–O bending and Ca–O stretching vibrations, respectively. These assignments are in good agreement with the functional groups of bredigite (Khandan and Ozada, 2017; Mirhadi et al., 2012; Rahmati et al., 2018). For the vancomycin-loaded scaffolds (Figures 4b and 4c), the broad peak of 3406 cm$^{-1}$ is attributed to the stretching vibration of the hydroxyl group. Also, peaks at 1655, 1504 and 1231 cm$^{-1}$ are assigned to the vibrational modes of C=O, aromatic C=C and phenolic cycle, respectively (Cerchiara et al., 2015; Ordikhani and Simchi, 2014; Yao et al., 2013), proving the successful incorporation of vancomycin into the scaffolds. The lack of any shifts in the absorption bands of vancomycin suggests that no chemical interactions occur between the drug and other substances, as an evidence for the physical adsorption of the drug. Also, for the PLGA-coated sample (Figure 4c), the characteristic peaks of PLGA are identified at 1182 and 1089 cm$^{-1}$ (C–O–C stretching), 1752 cm$^{-1}$ (C=O stretching) and 1454 cm$^{-1}$ (C–H stretching), which are compatible with the FTIR assignments of Refs. (Akl et al., 2016; Meng et al., 2010; Singh et al., 2014; Wang et al., 2019) for this aliphatic polyester.

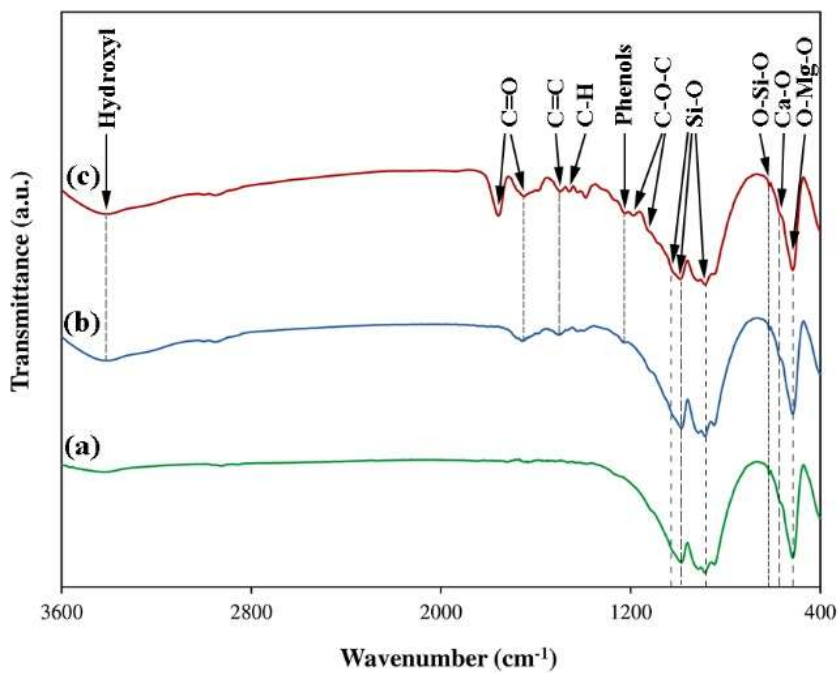

Figure 4. FTIR spectra of the bare vancomycin-free (a), bare vancomycin-loaded (b) and PLGA-coated vancomycin-loaded (c) bredigite scaffolds.

### *3.2. Drug delivery assessments*

The accumulative amount and percentage profiles of vancomycin released from the scaffolds are illustrated in Figure 5. The percentages were calculated with respect to the total loaded content of vancomycin which was 66.6±2.1% of the drug dissolved in the solution. For the bare scaffold, a burst release of vancomycin is detected within 9 h, so that about 94.7±2.5% of the loaded vancomycin level is released. Nevertheless, after coating of the scaffolds with PLGA, the burst release is beneficially declined to the first 6 h with only 21.0±2.9 and 18.5±1.8% of the loaded drug amount for the bredigite-5% PLGA and bredigite-10% PLGA scaffolds, respectively. These modified levels of the burst release in the first 6 h of implantation, called "decisive period", are essential to prevent the adhesion of bacteria and thereby inhibit infections (Linke and Goldman, 2011; Poelstra et al., 2002).

In accordance with Figure 5, after the initial burst release of the drug from the scaffolds, the release rate of vancomycin decreases and a sustained release is followed for the coated scaffolds. Typically, after one day of exposure, vancomycin is almost completely released from the bare scaffold, which is insufficient for the antibiotic therapy of osteomyelitis (Ford and Cassat, 2017; Long et al., 2016; Pande et al., 2015). In a similar study, Yao et al. (Yao et al., 2013) investigated the vancomycin release behavior of bare and coated bioglass scaffolds and showed that vancomycin was released completely from the bare scaffold in less than 24 h. In contrast, only 33.3±4.6 and 29.6±2.6% of the initial loaded drug amount are desirably released from the 5 and 10 % PLGA-coated bredigite scaffolds, respectively. Typically, after 7 days of immersion, nearly 6.16±2.2 and 52.77±2.7% of the drug still remain in the bredigite-5 % PLGA and bredigite-10 % PLGA scaffolds, respectively. The different kinetics of vancomycin release from the PLGA-coated samples is attributed to the different encapsulation modes of the struts with the PLGA deposits (Figures 2d and 2f) and the different thicknesses of the coatings (Figure 3). It is also worth mentioning that the drug release contents from the PLGA-coated scaffolds over the whole study period are above the MIC and MBC levels of vancomycin against *staphylococcus aureus* (0.75~2 and 8 μg/ml, respectively (Coudron et al., 1987; Mason et al., 2009)), which is desirable for treating osteomyelitis.

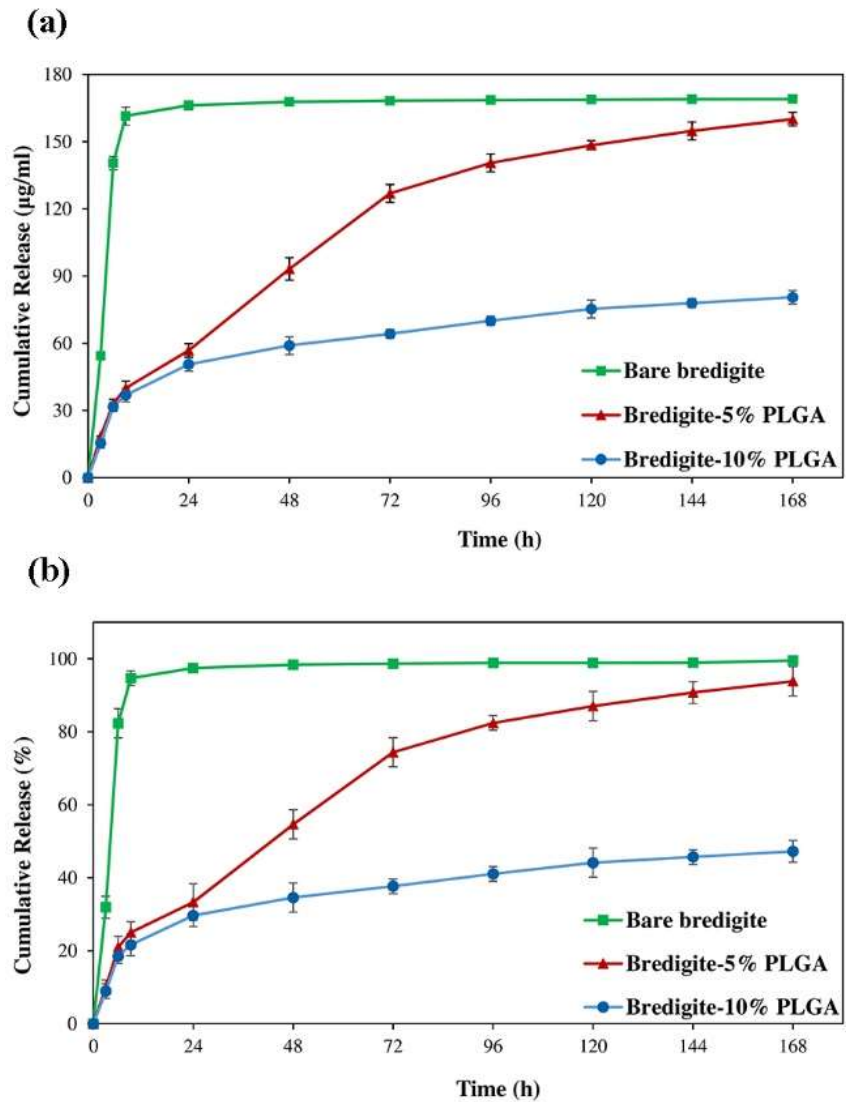

Figure 5. Concentration (a) and percentage (b) of vancomycin released from the different scaffolds.

### 3.3. Investigation of physiological pH variations

Figure 6 depicts the pH value of the SBF during exposure to the scaffolds as a biodegradation/bioresorption-related evaluation. A sharp increase of pH from 7.47±0.02 to 9.17±0.08 is observed for the bare bredigite scaffold during 7 days of soaking. Due to the domination of Ca in the bredigite structure ($Ca_7MgSi_4O_{16}$) and the high affinity of the Ca-O bond for hydrolysis, the bioresorption rate of bredigite is very high (Keihan and Salahinejad, 2020; Wu and Chang, 2007a; Zirak et al., 2020). This results in a high local concentration of $Ca^{+2}$ and $Mg^{+2}$ ions in the surrounding biological environment. Accordingly, the activation of the ion exchange mechanism reduces the hydrogen ion concentration of the environment and thereby enhances pH. Regarding the bare scaffold impregnated with the drug, the presence of vancomycin despite an acidic character (Marshall, 1965) could not significantly decline physiological pH, due to the low concentration of the drug released into the medium. In contrast, the PLGA coating of the vancomycin-loaded scaffolds effectively considerably buffers pH. According to Figure 6, the buffering effect is enhanced by increasing the PLGA concentration from 5% to 10 % due to the following reasons:

i. the release rate of the cations from the bredigite scaffolds is somewhat restricted with the increase of the coating thickness, where the PLGA coatings act as a physical barrier.

ii. the amount of the acidic coproducts of PLGA (i.e. glycolic and lactic acids) released due to hydrolytic biodegradation is enhanced by increasing the content of PLGA.

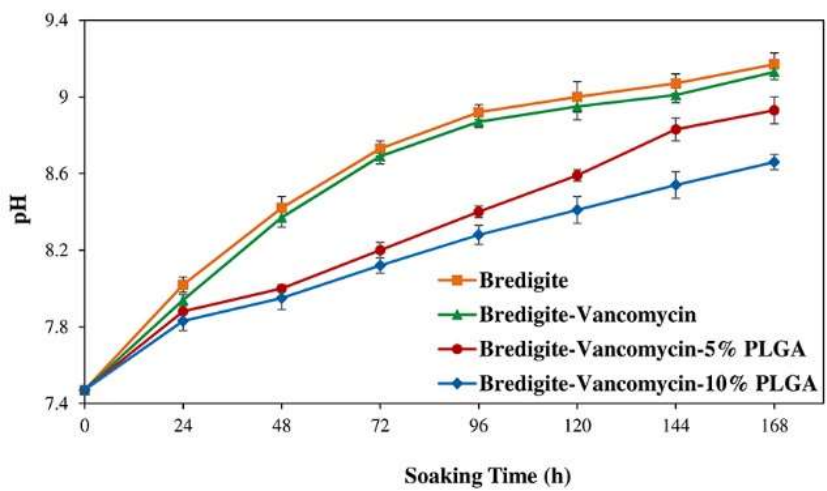

Figure 6. pH variations of the SBF in contact with the different scaffolds.

## *3.4. Biocompatibility evaluations*

The MTT assay results of the cells cultured on the different scaffolds are indicated in Figure 7, with the significance level of less than 5% (p-value<0.05). Over the entire periods of culture, the cytocompatibility of the samples with respect to dental pulp stem cells ranks as follows: vancomycin-loaded 10 % PLGA-coated bredigite > vancomycin-loaded 5 % PLGA-coated > bredigite > vancomycin-loaded bredigite. Essentially, the fast bioresorption of bredigite leads to high physiological pH (Figure 6) and the high concentration of $Ca^{2+}$ released into the medium (Peshwa et al., 1993; Zhivotovsky and Orrenius, 2011), which are responsible for the relatively low cell viability on the bredigite scaffold, in agreement with Ref. (Wu and Chang, 2007a). The impregnation of the bredigite scaffold with the antibiotic drug under the bare conditions further decreases the cell cytocompatibility, due to the burst release of vancomycin in the medium (Figure 5). Consistently, the detrimental influence of high vancomycin dosages on biocompatibility has been previously pointed out (Bakhsheshi-Rad et al., 2017). The controlling effects of the PLGA coatings on the burst release of the drug (Figure 5), bredigite bioresorption and physiological pH (Figure 6) cause the higher cytocompatibility of the coated scaffolds. This is supported by the fact that the cell viability is further improved by increasing the PLGA concentration from 5% to 10%.

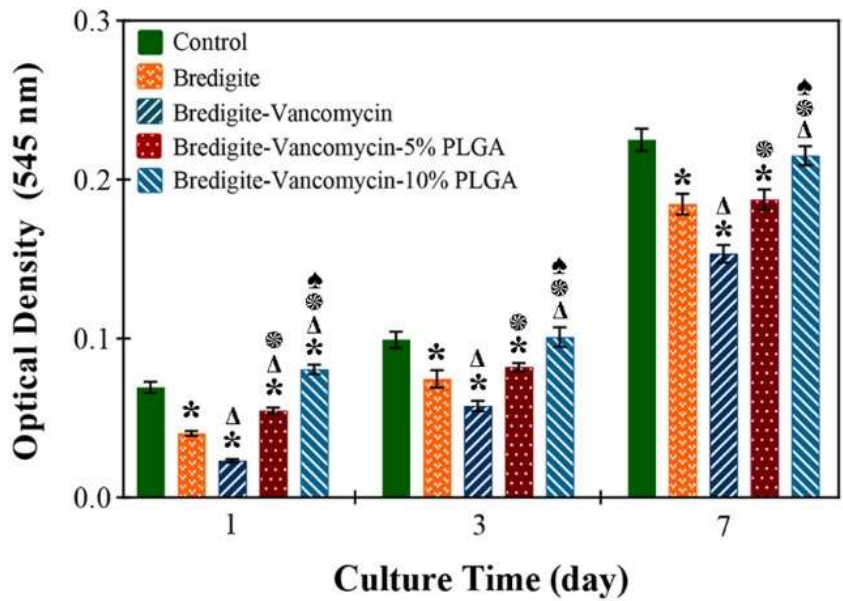

Figure 7. MTT results of the cell cultures on the different scaffolds, where *, Δ, ✿ and ♠ demonstrate significant differences with respect to the control, bare vancomycin-free bredigite, vancomycin-loaded bredigite and vancomycin-loaded 5% PLGA-coated bredigite scaffolds, respectively.

Form this work, it is finally concluded that the 10% PLGA coating is the optimal surface modification for the vancomycin-loaded bredigite scaffolds for bone tissue-engineering and local drug delivery applications. This assignment is due to the fact that this sample benefits from the appropriate pores configuration, drug release kinetics, cytocompatibility and cell attachment characterized by a spread morphology of cells with high cytoplasmic extensions (Figure 8).

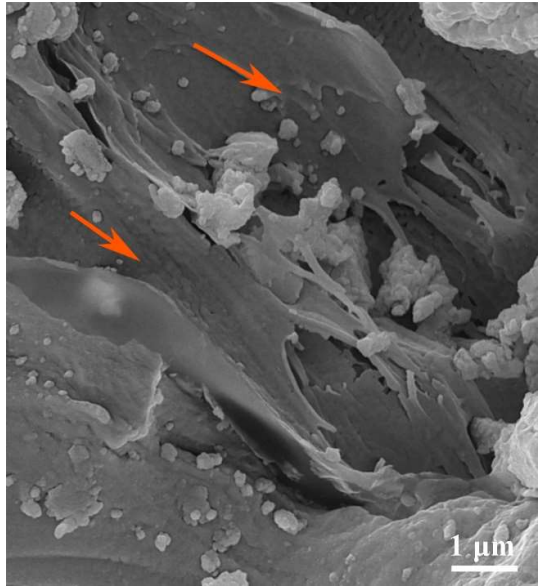

Figure 8. FESEM micrograph of a cell cultured on the vancomycin-impregnated 10% PLGA-coated bredigite scaffold, in which the arrows indicate cells.

## 4. Conclusions

In this work, bredigite ($Ca_7MgSi_4O_{16}$) porous scaffolds were fabricated by the sponge replica method, impregnated with vancomycin hydrochloride and coated with PLGA for bone tissue engineering. The bare drug-loaded scaffold exhibited a one-step burst release of vancomycin, so that the relatively entire amount of the drug loaded was released in less than 24 h. The burst release of the drug and the fast bioresorption of bredigite were recognized to be responsible for the relatively low cytocompatibility of this sample with respect to dental pulp stem cells. The PLGA coating, nevertheless, altered the drug delivery kinetics characterized by the limitation of the burst drug release stage and by the development of a sustained drug release behavior. This surface modification process also improved the cell viability, which was attributed to the modification of the drug release and the buffering effect of PLGA on physiological pH. The most improvement in the biological behaviors was obtained by employing the PLGA coating deposited from the 10% PLGA-acetone solution, with the suitable characteristics of pores for bone tissue engineering and the appropriate antibiotic release kinetics for osteomyelitis treatment.

# References

Akl, M.A., Kartal-Hodzic, A., Oksanen, T., Ismael, H.R., Afouna, M.M., Yliperttula, M., Samy, A.M., Viitala, T., 2016. Factorial design formulation optimization and in vitro characterization of curcumin-loaded PLGA nanoparticles for colon delivery. Journal of Drug Delivery Science and Technology 32, 10-20.

Antoci Jr, V., Adams, C.S., Hickok, N.J., Shapiro, I.M., Parvizi, J., 2007. Antibiotics for local delivery systems cause skeletal cell toxicity in vitro. Clinical Orthopaedics and Related Research 462, 200-206.

Bakhsheshi-Rad, H., Hamzah, E., Ismail, A., Aziz, M., Hadisi, Z., Kashefian, M., Najafinezhad, A., 2017. Novel nanostructured baghdadite-vancomycin scaffolds: In-vitro drug release, antibacterial activity and biocompatibility. Materials Letters 209, 369-372.

Bose, S., Sarkar, N., Banerjee, D., 2018. Effects of PCL, PEG and PLGA polymers on curcumin release from calcium phosphate matrix for in vitro and in vivo bone regeneration. Materials today chemistry 8, 110-120.

Cerchiara, T., Abruzzo, A., Di Cagno, M., Bigucci, F., Bauer-Brandl, A., Parolin, C., Vitali, B., Gallucci, M., Luppi, B., 2015. Chitosan based micro-and nanoparticles for colon-targeted delivery of vancomycin prepared by alternative processing methods. European Journal of Pharmaceutics and Biopharmaceutics 92, 112-119.

Coudron, P.E., Johnston, J.L., Archer, G.L., 1987. In-vitro activity of LY146032 against Staphylococcus aureus and S. epidermidis. Journal of Antimicrobial Chemotherapy 20, 505-511.

Diba, M., Goudouri, O.-M., Tapia, F., Boccaccini, A.R., 2014. Magnesium-containing bioactive polycrystalline silicate-based ceramics and glass-ceramics for biomedical applications. Current Opinion in Solid State and Materials Science 18, 147-167.

Diba, M., Tapia, F., Boccaccini, A.R., Strobel, L.A., 2012. Magnesium-containing bioactive glasses for biomedical applications. International Journal of Applied Glass Science 3, 221-253.

Edin, M.L., Miclau, T., Lester, G.E., Lindsey, R.W., Dahners, L.E., 1996. Effect of cefazolin and vancomycin on osteoblasts in vitro. Clinical Orthopaedics and Related Research 333, 245-251.

Ferguson, J., Dudareva, M., Riley, N., Stubbs, D., Atkins, B., McNally, M., 2014. The use of a biodegradable antibiotic-loaded calcium sulphate carrier containing tobramycin for the treatment of chronic osteomyelitis: a series of 195 cases. The Bone & Joint journal 96, 829-836.

Ford, C.A., Cassat, J.E., 2017. Advances in the local and targeted delivery of anti-infective agents for management of osteomyelitis. Expert review of anti-infective therapy 15, 851-860.

Galla, J.H., 2000. Metabolic alkalosis. Journal of the American Society of Nephrology 11, 369-375.

Gold, H.S., Moellering Jr, R.C., 1996. Antimicrobial-drug resistance. New England Journal of Medicine 335, 1445-1453.

Hetrick, E.M., Schoenfisch, M.H., 2006. Reducing implant-related infections: active release strategies. Chemical Society Reviews 35, 780-789.

Huang, X., Brazel, C.S., 2001. On the importance and mechanisms of burst release in matrix-controlled drug delivery systems. Journal of Controlled Release 73, 121-136.

Jadidi, A., Salahinejad, E., 2020. Mechanical strength and biocompatibility of bredigite (Ca7MgSi4O16) tissue-engineering scaffolds modified by aliphatic polyester coatings. Ceramics International 46, 16439–16446.

Jain, K.K., 2008. Drug delivery systems-an overview, Drug delivery systems. Springer, pp. 1-50.

Keihan, R., Ghorbani, A., Salahinejad, E., Sharifi, E., Tayebi, L., 2020. Biomineralization, strength and cytocompatibility improvement of bredigite scaffolds through doping/coating. Ceramics International 46, 21056–21063.

Keihan, R., Salahinejad, E., 2020. Inorganic-salt coprecipitation synthesis, fluoride-doping, bioactivity and physiological pH buffering evaluations of bredigite. Ceramics International 46, 13292-13296.

Khandan, A., Ozada, N., 2017. Bredigite-Magnetite (Ca7MgSi4O16-Fe3O4) nanoparticles: A study on their magnetic properties. Journal of Alloys and Compounds 726, 729-736.

Khojasteh, A., Fahimipour, F., Eslaminejad, M.B., Jafarian, M., Jahangir, S., Bastami, F., Tahriri, M., Karkhaneh, A., Tayebi, L., 2016. Development of PLGA-coated β-TCP scaffolds containing VEGF for bone tissue engineering. Materials Science and Engineering: C 69, 780-788.

Krishnan, A.G., Jayaram, L., Biswas, R., Nair, M., 2015. Evaluation of antibacterial activity and cytocompatibility of ciprofloxacin loaded Gelatin–Hydroxyapatite scaffolds as a local drug delivery system for osteomyelitis treatment. Tissue Engineering Part A 21, 1422-1431.

Lee, S., Porter, M., Wasko, S., Lau, G., Chen, P.-Y., Novitskaya, E.E., Tomsia, A.P., Almutairi, A., Meyers, M.A., McKittrick, J., 2012. Potential bone replacement materials prepared by two methods. MRS Online Proceedings Library Archive 1418.

Linke, D., Goldman, A., 2011. Bacterial adhesion: chemistry, biology and physics. Springer Science & Business Media.

Long, Y., Zhu, Z., Yu, Y., Zhang, S., 2016. Progress of different drug delivery route of vancomycin for the treatment of chronic osteomyelitis. Zhonghua wai ke za zhi [Chinese journal of surgery] 54, 716-720.

Lucke, M., Schmidmaier, G., Sadoni, S., Wildemann, B., Schiller, R., Stemberger, A., Haas, N., Raschke, M., 2003. A new model of implant-related osteomyelitis in rats. Journal of Biomedical Materials Research Part B: Applied Biomaterials 67, 593-602.

Marshall, F.J., 1965. Structure studies on vancomycin. Journal of medicinal chemistry 8, 18-22.

Marwah, H., Garg, T., Goyal, A.K., Rath, G., 2016. Permeation enhancer strategies in transdermal drug delivery. Drug delivery 23, 564-578.

Mason, E.O., Lamberth, L.B., Hammerman, W.A., Hulten, K.G., Versalovic, J., Kaplan, S.L., 2009. Vancomycin MICs for Staphylococcus aureus vary by detection method and have subtly increased in a pediatric population since 2005. Journal of Clinical Microbiology 47, 1628-1630.

Maurmann, N., Pereira, D.P., Burguez, D., de S Pereira, F.D., Neto, P.I., Rezende, R.A., Gamba, D., da Silva, J.V., Pranke, P., 2017. Mesenchymal stem cells cultivated on scaffolds formed by 3D printed PCL matrices, coated with PLGA electrospun nanofibers for use in tissue engineering. Biomedical Physics & Engineering Express 3, 045005.

Meng, Z., Wang, Y., Ma, C., Zheng, W., Li, L., Zheng, Y., 2010. Electrospinning of PLGA/gelatin randomly-oriented and aligned nanofibers as potential scaffold in tissue engineering. Materials Science and Engineering: C 30, 1204-1210.

Mirhadi, S., Tavangarian, F., Emadi, R., 2012. Synthesis, characterization and formation mechanism of single-phase nanostructure bredigite powder. Materials Science and Engineering: C 32, 133-139.

This is the accepted manuscript (postprint) of the following article:
A. Jadidi, E. Salahinejad, E. Sharifi, L. Tayebi, *Drug-delivery Ca-Mg silicate scaffolds encapsulated in PLGA*, International Journal of Pharmaceutics, 589 (2020) 119855.
https://doi.org/10.1016/j.ijpharm.2020.119855
Olalde, B., Garmendia, N., Sáez-Martínez, V., Argarate, N., Nooeaid, P., Morin, F., Boccaccini, A., 2013. Multifunctional bioactive glass scaffolds coated with layers of poly (d, l-lactide-co-glycolide) and poly (n-isopropylacrylamide-co-acrylic acid) microgels loaded with vancomycin. Materials Science and Engineering: C 33, 3760-3767.

Ordikhani, F., Simchi, A., 2014. Long-term antibiotic delivery by chitosan-based composite coatings with bone regenerative potential. Applied Surface Science 317, 56-66.

Pande, K.C., Putera, J., Seri, B., 2015. Optimal management of chronic osteomyelitis: current perspectives. Orthopedic Research and Reviews 7, 71-81.

Park, S.H., Park, D.S., Shin, J.W., Kang, Y.G., Kim, H.K., Yoon, T.R., Shin, J.-W., 2012. Scaffolds for bone tissue engineering fabricated from two different materials by the rapid prototyping technique: PCL versus PLGA. Journal of Materials Science: Materials in Medicine 23, 2671-2678.

Peshwa, M.V., Kyung, Y.S., McClure, D.B., Hu, W.S., 1993. Cultivation of mammalian cells as aggregates in bioreactors: effect of calcium concentration of spatial distribution of viability. Biotechnology and bioengineering 41, 179-187.

Poelstra, K.A., Barekzi, N.A., Rediske, A.M., Felts, A.G., Slunt, J.B., Grainger, D.W., 2002. Prophylactic treatment of gram-positive and gram-negative abdominal implant infections using locally delivered polyclonal antibodies. Journal of biomedical materials research 60, 206-215.

Rahmati, M., Fathi, M., Ahmadian, M., 2018. Preparation and structural characterization of bioactive bredigite ($Ca_7MgSi_4O_{16}$) nanopowder. Journal of Alloys and Compounds 732, 9-15.

Sankar, V., Hearnden, V., Hull, K., Juras, D.V., Greenberg, M., Kerr, A., Lockhart, P.B., Patton, L.L., Porter, S., Thornhill, M., 2011. Local drug delivery for oral mucosal diseases: challenges and opportunities. Oral diseases 17, 73-84.

Schumacher, T.C., Volkmann, E., Yilmaz, R., Wolf, A., Treccani, L., Rezwan, K., 2014. Mechanical evaluation of calcium-zirconium-silicate (baghdadite) obtained by a direct solid-state synthesis route. Journal of the Mechanical Behavior of Biomedical Materials 34, 294-301.

Shahrouzifar, M., Salahinejad, E., 2019. Strontium doping into diopside tissue engineering scaffolds. Ceramics International 45, 10176-10181.

Shahrouzifar, M., Salahinejad, E., Sharifi, E., 2019. Co-incorporation of strontium and fluorine into diopside scaffolds: Bioactivity, biodegradation and cytocompatibility evaluations. Materials Science and Engineering: C 103, 109752.

Shirtliff, M., Mader, J., 2000. Osteomyelitis, Persistent bacterial infections. American Society of Microbiology, pp. 375-396.

Singh, G., Kaur, T., Kaur, R., Kaur, A., 2014. Recent biomedical applications and patents on biodegradable polymer-PLGA. International Journal of Pharmacy and Pharmaceutical Sciences 1, 30-32.

Soundrapandian, C., Datta, S., Kundu, B., Basu, D., Sa, B., 2010. Porous bioactive glass scaffolds for local drug delivery in osteomyelitis: development and in vitro characterization. AAPS Pharmscitech 11, 1675-1683.

Su, W., Hu, Y., Zeng, M., Li, M., Lin, S., Zhou, Y., Xie, J., 2019. Design and evaluation of nano-hydroxyapatite/poly (vinyl alcohol) hydrogels coated with poly (lactic-co-glycolic acid)/nano-hydroxyapatite/poly (vinyl alcohol) scaffolds for cartilage repair. Journal of Orthopaedic Surgery and Research 14, 446.

Thomas, M.J.K., 1996. Ultraviolet & Visible Spectroscopy. John Wiley & Sons.

Wang, J., Helder, L., Shao, J., Jansen, J.A., Yang, M., Yang, F., 2019. Encapsulation and release of doxycycline from electrospray-generated PLGA microspheres: Effect of polymer end groups. International Journal of Pharmaceutics 564, 1-9.

Weiner, S., Wagner, H.D., 1998. The material bone: structure-mechanical function relations. Annual Review of Materials Science 28, 271-298.
24

Wu, C., Chang, J., 2007a. Degradation, bioactivity, and cytocompatibility of diopside, akermanite, and bredigite ceramics. Journal of Biomedical Materials Research Part B: Applied Biomaterials 83, 153-160.

Wu, C., Chang, J., 2007b. Synthesis and in vitro bioactivity of bredigite powders. Journal of Biomaterials Applications 21, 251-263.

Wu, C., Chang, J., Zhai, W., Ni, S., 2007. A novel bioactive porous bredigite (Ca 7 MgSi 4 O 16) scaffold with biomimetic apatite layer for bone tissue engineering. Journal of Materials Science: Materials in Medicine 18, 857-864.

Wu, C., Chang, J., Zhai, W., Ni, S., Wang, J., 2006. Porous akermanite scaffolds for bone tissue engineering: preparation, characterization, and in vitro studies. Journal of Biomedical Materials Research Part B: Applied Biomaterials: An Official Journal of The Society for Biomaterials, The Japanese Society for Biomaterials, and The Australian Society for Biomaterials and the Korean Society for Biomaterials 78, 47-55.

Wu, C., Luo, Y., Cuniberti, G., Xiao, Y., Gelinsky, M., 2011. Three-dimensional printing of hierarchical and tough mesoporous bioactive glass scaffolds with a controllable pore architecture, excellent mechanical strength and mineralization ability. Acta biomaterialia 7, 2644-2650.

Wu, C., Ramaswamy, Y., Zreiqat, H., 2010. Porous diopside (CaMgSi2O6) scaffold: a promising bioactive material for bone tissue engineering. Acta biomaterialia 6, 2237-2245.

Yao, Q., Nooeaid, P., Roether, J.A., Dong, Y., Zhang, Q., Boccaccini, A.R., 2013. Bioglass®-based scaffolds incorporating polycaprolactone and chitosan coatings for controlled vancomycin delivery. Ceramics International 39, 7517-7522.

Zamboni, F., Keays, M., Hayes, S., Albadarin, A.B., Walker, G.M., Kiely, P.A., Collins, M.N., 2017. Enhanced cell viability in hyaluronic acid coated poly (lactic-co-glycolic acid) porous scaffolds within microfluidic channels. International Journal of Pharmaceutics 532, 595-602.

Zhao, L., Lin, K., Zhang, M., Xiong, C., Bao, Y., Pang, X., Chang, J., 2011. The influences of poly (lactic-co-glycolic acid)(PLGA) coating on the biodegradability, bioactivity, and biocompatibility of calcium silicate bioceramics. Journal of Materials Science 46, 4986-4993.

Zhao, L., Wu, C., Lin, K., Chang, J., 2012. The effect of poly (lactic-co-glycolic acid)(PLGA) coating on the mechanical, biodegradable, bioactive properties and drug release of porous calcium silicate scaffolds. Bio-medical Materials and Engineering 22, 289-300.

Zhivotovsky, B., Orrenius, S., 2011. Calcium and cell death mechanisms: a perspective from the cell death community. Cell calcium 50, 211-221.

Zirak, N., Jahromi, A.B., Salahinejad, E., 2020. Vancomycin release kinetics from Mg–Ca silicate porous microspheres developed for controlled drug delivery. Ceramics International 46, 508-512.